\documentclass[%
aps,
 prl,%
 amsmath,amssymb,showpacs,
 reprint,%
twocolumn
]{revtex4-1}

\usepackage{graphicx}
\usepackage{dcolumn}
\usepackage{bm}

\usepackage{color}

\newcommand{\PR }{Phys. Rev. }
\newcommand{\PRL }{Phys. Rev. Lett. }
\newcommand{\PRA }{\PR A }

\newcommand{\NJP }{New J. Phys. }
\newcommand{\Eqref}[1]{\mbox{Eq.~(\ref{#1})}}
\newcommand{\figref}[1]{\mbox{Fig.~\ref{#1}}}

\newcommand{\vv}[1]{\mathbf{#1}}         
\newcommand{\mathperiod}{\,.} 
\newcommand{\mathcomma}{\,,}  
\newcommand{\ee}{\mathrm{e}}             
\newcommand{\ii}{\mathrm{i}}             
\newcommand{\abs}[1]{\left|{#1}\right|}
\newcommand{\abssmall}[1]{|{#1}|}

\newcommand{\ketsmall}[1]{|{#1}\rangle}

\newcommand{\brasmall}[1]{\langle{#1}|}
\newcommand{\expvalsmall}[1]{\langle{#1}\rangle}
\newcommand{\ie}{\mbox{i.\,e.}\nolinebreak[4]}
\newcommand{\eg}{\mbox{e.\,g.}\nolinebreak[4]}
\newcommand{\cf}{{cf.}~}
\newcommand{\anfz}[1]{``{#1}''}

\begin{document}

\title{Far-Field Signatures of a Two-Body Bound State 
	in Collective Emission from Interacting 
	Two-Level Atoms on a Lattice}

\author{Paolo~Longo}
\email{paolo.longo@mpi-hd.mpg.de}
\author{J\"org~Evers}%
\affiliation{Max Planck Institute for Nuclear Physics, Saupfercheckweg 1, 69117 Heidelberg, Germany}

\date{\today}

\begin{abstract}
{The collective emission from a one-dimensional chain of interacting two-level 
atoms is investigated.
We calculate 
the light scattered by dissipative few-excitation eigenstates in the far-field, 
and in particular focus on signatures of a lattice two-body bound state.
We present analytical results for the angle-resolved, temporal decay of 
the scattered light intensity. Moreover, we find that the steady-state 
emission spectrum that emerges when the system is probed by a weak, incoherent 
driving field exhibits a distinct signature for the existence of a bound state, 
and allows to determine the momentum distribution of the two-body relative wavefunction. 
Intriguingly, our study does not rely on single-atom addressability and/or 
manipulation techniques.
}
\end{abstract}

\pacs{42.50.Nn, 42.50.Ct}


\maketitle

Over the last years, artificially designed lattice systems
have become the focus of intense
experimental and theoretical research across various subdisciplines 
in quantum optics. 
Realizations that have already reached a highly sophisticated level 
of control over basic mechanisms of light--matter interaction
include, to name just a few, 
cold atoms in optical lattices \cite{zoller05,winkler06,fukuhara13}, 
fiber-based settings \cite{rauschenbeutel10}, 
atom--cavity networks \cite{angelakis08,koch13}, 
or on-chip photonics \cite{benson13}.

Despite these promising avenues,
long-standing, fundamental questions are far from being outdated and 
are just at the edge of what is realizable experimentally today.
For instance, while the prediction of a two-body bound state on a lattice 
dates back to Bethe \cite{bethe32}, experimental investigations have become
possible only recently in the context of the Bose-Hubbard model \cite{winkler06} or 
the Heisenberg model \cite{fukuhara13}. In line with these experimental advancements,
the scope of recent theoretical studies includes the study of 
the few-excitation eigenstates
of 1d lattice systems \cite{valiente08,valiente09,longo13}, 
aspects of dynamics \cite{evertz12,longo13},
or issues of entanglement and coherence \cite{porras13}.
However, the recent experiments on the two-body bound state~\cite{winkler06,fukuhara13} 
are demanding in that they require {\it in situ} tuning of parameters and/or single-site manipulation. 
This prompts for alternative approaches. 
A promising candidate is the coupling to a probing light field~\cite{Weitenberg11,porras08,lesanovsky10,Mekhov12}, 
which raises the question of how much information about an exotic two-body bound state 
on a lattice can be inferred from the optical far field.

In this paper, we analytically calculate the scattered light from a 1d lattice of atoms, 
and show that a dissipative and collective two-body bound state 
imprints an unambiguous characteristic far-field signature onto the light. 
Unlike in recent state-of-the art cold-atom experiments \cite{winkler06,fukuhara13},
the bound state discussed in the present work is like a molecule 
of two quanta stored as atomic excitation in a lattice of immobile 
atoms, rather than a composite object of two atoms tunneling in an 
optical lattice.
We employ a description reminiscent of spin physics that 
allows us to discuss the relevant physical mechanisms in a broader 
context. Relating atomic operators to the emitted light field and based on 
the dissipative dynamics as described by a Lindblad equation, 
we utilize the Glauber decomposition \cite{glauber} to obtain 
far-field observables that are amenable to standard experimental 
techniques.
We first identify the fingerprint for the two-body bound state in its
spontaneous emission dynamics. Then, we show that the same distinct feature 
is also present in the emission spectrum when the system is probed 
by a weak, incoherent driving field.
Furthermore, we discuss a method to infer the bound state's
momentum distribution from the far field.
These findings represent a simple means for the identification of a 
two-body bound state without the need to individually address and 
manipulate single atoms. 
From a broader perspective,
our investigations open up an alternative approach for the study of exotic 
excitations in a quantum-optical context.

To begin with, we introduce our model based on 
its underlying building block, which is a single two-level atom.
Being coupled to an electromagnetic reservoir (\eg, free space), 
the bare atomic transition frequency $\omega_0$
is shifted by $\mathrm{Im}(\Gamma_0)/2=\gamma_0/\pi$ (Lamb shift) and
the atom is subject to spontaneous decay with a rate of $\mathrm{Re}(\Gamma_0)=\gamma_0$
\cite{li12}.
If we imagine an identical, second atom nearby, 
photons can be exchanged between the two atoms
via the common electromagnetic reservoir by virtue of dipole--dipole coupling.
Let the amplitude of this process be proportional to $\mathrm{Im}(\Gamma_1)$
and the dissipative part (\ie, irreversible photon loss to the reservoir) 
be characterized by $\mathrm{Re}(\Gamma_1)$. 
To give one example, 
for atoms embedded in free space with dipole moments aligned perpendicular to the inter-atomic distance vector
(magnitude $a$),
we have \cite{li12}
$\Gamma_1 = - (3 \ii \gamma_0 \lambda_{\mathrm{at}}/4 \pi a) \exp(2 \pi \ii a / \lambda_{\mathrm{at}})$,
where $\lambda_{\mathrm{at}}= 2\pi c / \omega_0$ (speed of light $c$).
Besides these dissipative aspects, atom--atom interactions 
result in an energy shift $U$ if both atoms are in the excited state.

\begin{figure}[t]
 \centering
  \includegraphics[width=0.46\textwidth]{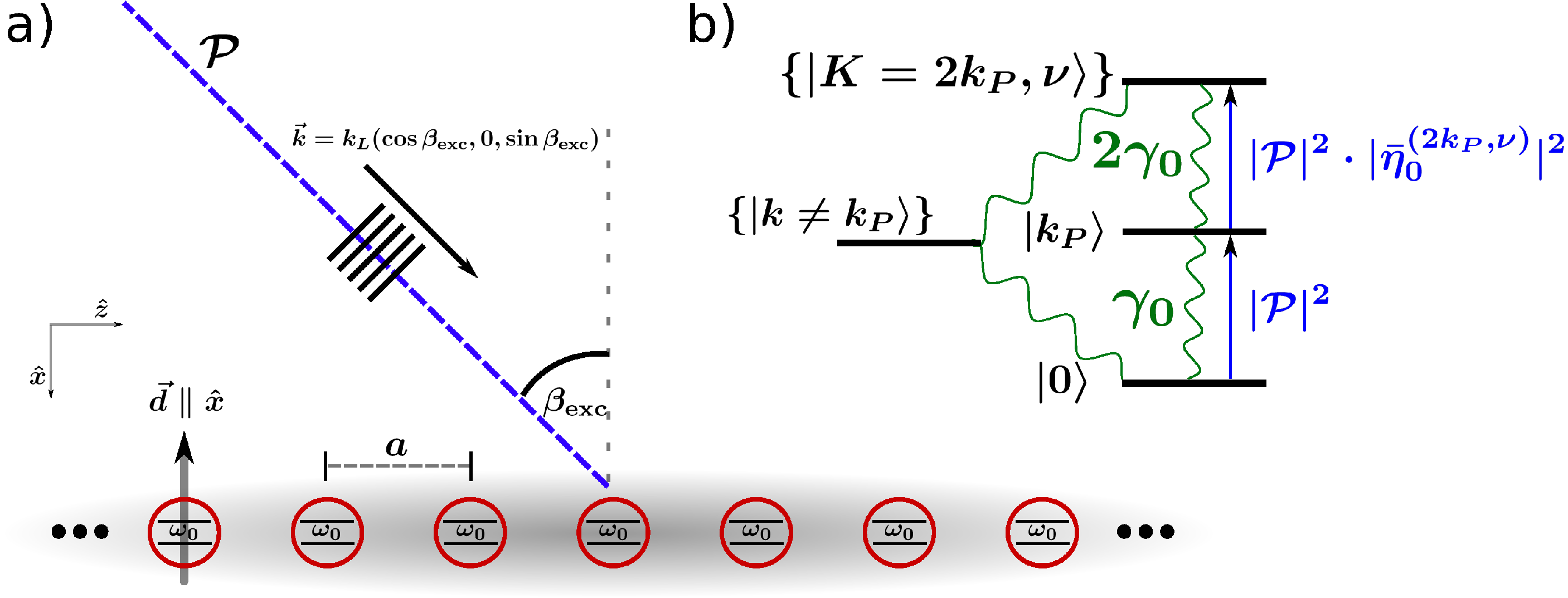}
  \caption{\label{fig:sketchandredHS}
	    a) 
	    1d lattice of two-level atoms (dipole moments aligned 
	    parallel to the $x$-axis).
	    b) 
	    An incoherent drive 
	    (pump rate $\abssmall{\mathcal{P}}^2$, 
	    angle $\beta_{\mathrm{exc}} \neq 0$) 
	    \anfz{imprints} the wavenumber~$k_P$ 
	    so that the relevant Hilbert space comprises only two-excitation 
	    states with $K=2 k_P$.
	    The pump's electric field polarization vector lies in the $x$-$z$-plane and
	    is perpendicular to $\vv{k}$.
	    }
\end{figure}

Combining these elements, we can construct the Hamiltonian for 
a 1d lattice of 
$M \gg 1$ atoms (see also sketch in \figref{fig:sketchandredHS}a)).
In this paper, we focus on a situation where the bare atomic emission wavelength is smaller
than the lattice constant, \ie, $\lambda_{\mathrm{at}}/a < 1$, realizing the 
\anfz{extended-sample regime} 
(contrasting the established \anfz{small-volume limit} \cite{dicke54,haroche82}).
In this regime, we can restrict the dipole--dipole coupling and the atom--atom interactions 
to nearest neighbors.
Based on Ref.~\cite{li12}, we formulate the 
Hamiltonian 
($\hbar \equiv 1$, $M \equiv N+1$ is odd)
\begin{eqnarray}
 \label{eq:theH}
 && \hat{H} = 
  \sum_{n=-N/2}^{N/2} \left( \omega_0 - \frac{\ii \Gamma_0}{2} \right) \sigma^+_{n} \sigma^-_n
  \\
 \nonumber
 && ~~~
  + \sum_{n=-N/2}^{N/2-1}
  \left[
	- \frac{\ii \Gamma_1}{2} \left( \sigma^+_{n+1} \sigma^-_n + \mathrm{h.c.} \right)
	+ U  \sigma^+_{n+1} \sigma^-_{n+1} \sigma^+_{n} \sigma^-_{n} 
  \right]
 \mathperiod
\end{eqnarray}
Here, 
$\sigma^+_{n}$ ($\sigma^-_{n}$) denotes a raising (lowering) operator for the atom at 
lattice site $n$ 
(satisfying the Pauli spin-$\frac{1}{2}$ operator algebra 
$\left[ \sigma^+_i, \sigma^-_j \right] = \left( 2 \sigma^+_i \sigma^-_i - 1 \right) \delta_{ij}$).
Since the rates $\Gamma_0$ and $\Gamma_1$ that enter this effective Hamiltonian 
are complex numbers,
the eigenproblem is non-Hermitian, which allows for the study of the 
dissipative and collective radiative eigenmodes.
Exemplarily, orders of magntiudes in the field of Rydberg atoms are \cite{weidemueller13}
$\lambda_{\mathrm{at}} \sim 500~\mathrm{nm}$ ($\omega_0 / 2 \pi \sim 500~\mathrm{THz}$), 
$\gamma_0 \sim \mathrm{MHz}$, and $U \sim 50~\mathrm{GHz}$ (for $a \sim 1~\mu\mathrm{m}$), 
representing separated scales $\omega_0 \gg U \gg \gamma_0$.
Also note that Hamiltonian~(\ref{eq:theH}) can be recast into the form of 
an $xyz$-type spin model (using the relation to the Pauli matrices 
$\sigma^{\pm}_n = \left( \sigma^x_n \pm \mathrm{i} \sigma^y_n \right) / 2$)
which finds application in the field of coupled-cavity arrays~\cite{plenio08}
or in the context of cold polar molecules~\cite{whaley14}.

In the subspace of a single excitation 
(\ie, $\mathcal{C} \equiv \sum_n \expvalsmall{\sigma^+_n \sigma^-_n} = 1$),
the interaction $U$ plays no role and 
the eigenstates $\ketsmall{k}=\sum_n \varphi^{(k)}_n \sigma^{+}_n \ketsmall{0}$ 
($\ketsmall{0}$ is the vacuum state)
are 1d spin waves
$\varphi^{(k)}_n = \exp\left( \ii k a n \right) / \sqrt{M}$
with a wavenumber 
$ka=-\pi + 2 \pi \ell / M$
from the first Brillouin zone ($\ell=0,\dots,M-1$).
The corresponding complex eigenenergy 
$E^{(1)}_k = \omega_0 -\ii \Gamma_0/2 -\ii \Gamma_1 \cos(ka)$
represents the dissipative dispersion relation of a tight-binding chain.
Assuming sharp optical transitions
($\gamma_0 \ll \omega_0$), 
we may neglect
the Lamb shift such that $\mathrm{Re}(E^{(1)}_k) \simeq \omega_0$.
For weak dipole--dipole coupling ($\lambda_{\mathrm{at}}/a < 1$),
$\mathrm{Im}(E^{(1)}_k) \simeq - \gamma_0/2$, 
which means \cite{li12}
that the 
single excitation's probability
decays at a rate~$\Gamma^{k}_0 \equiv -2 \mathrm{Im}(E^{(1)}_{k}) \simeq \gamma_0$.

The eigenstates for two excitations ($\mathcal{C}=2$)
can be written as $\ketsmall{K \nu} = \sum_{n_1 n_2} \Phi^{(K \nu)}_{n_1 n_2} \sigma^+_{n_1} \sigma^+_{n_2} \ketsmall{0}$,
where the two-spin wavefunction for $n_1 \neq n_2$,
$\Phi^{(K \nu)}_{n_1 n_2} = \exp\left[ \ii Ka \left(n_1+n_2\right)/2 \right] / (2\sqrt{M}) \cdot \Psi^{(K \nu)}_{n_1-n_2}$, 
is a product 
of a center-of-mass plane wave (wavenumber $K$) and a relative wavefunction
$\Psi^{(K \nu)}_{n_1-n_2}$ \cite{valiente08,valiente09,longo13,winkler06,fukuhara13}.
For $n_1 = n_2$, the wavefunction needs to vanish ($\Psi^{(K \nu)}_0 = 0$) 
since a single atom cannot be doubly excited,
expressing the fact that the excitations of a 1d spin-$\frac{1}{2}$ chain are hard-core bosons.
Originally put forward by Bethe \cite{bethe32} 
(and, for instance, also addresse in Refs.~\cite{valiente08,valiente09,longo13,winkler06,fukuhara13}),
is the remarkable fact that a complete basis of the two-excitation submanifold comprises
scattering states \emph{and} bound states.

Consequently, for each center-of-mass momentum $K$,
we have scattering states characterized by their
relative momentum $p$.
The relative wavefunction is of the form
$\Psi^{(K, \nu=p)}_{x \neq 0} \propto \exp \left( \ii p a \abs{x} \right) + \exp \left( - \ii p a \abs{x}  + \ii \delta_{Kp} \right) $,
where $\delta_{Kp}$ denotes the scattering phase shift induced by the interaction $U$
and describes the collision of two interacting spin waves.
Their complex eigenenergy is independent of the interaction strength $U$ and can be written as the sum
of single-excitation energies (as is always the case for scattering states), yielding
$E^{(2)}_{Kp} = E^{(1)}_{K/2+p} + E^{(1)}_{K/2-p} = 2 \omega_0 -\ii \Gamma_0 - 2\ii \Gamma_1 \cos(K/2) \cos{p}$.
As before, we may approximate this expression as 
$\mathrm{Re}(E^{(2)}_{Kp}) \simeq 2 \omega_0$
and
$\mathrm{Im}(E^{(2)}_{Kp}) \simeq -\gamma_0$.

\begin{figure}[t]
 \centering
\includegraphics[width=0.45\textwidth]{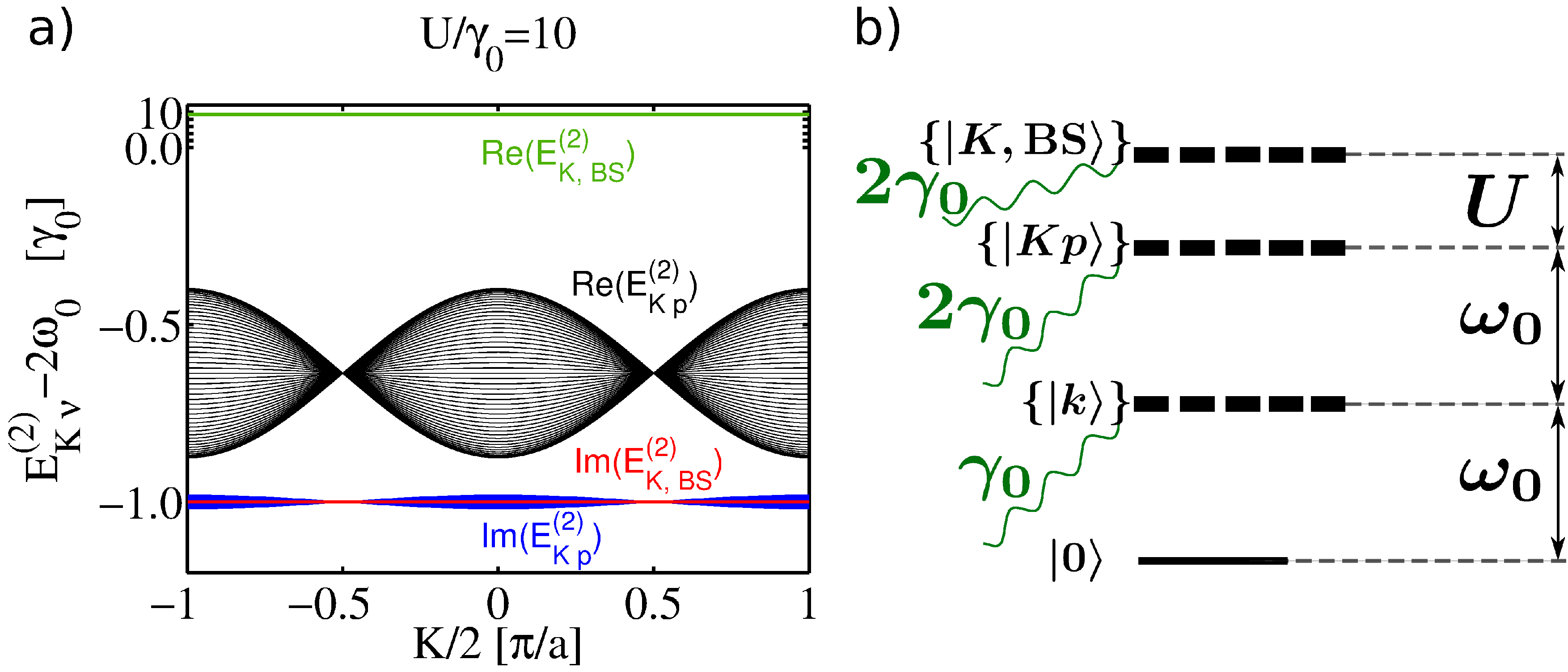}
  \caption{\label{fig:disprelandsimp}
	    a)
	    Complex two-excitation dispersion relation as a function of the center-of-mass momentum~$K$
	    ($\lambda_{\mathrm{at}}/a = 0.5$).
	    The different black and blue lines denote different relative momenta~$p$
	    for the scattering states.
	    The bound states' energies (real parts) are detached from the scattering states.
	    b)
	    Simplified level scheme for $\omega_0, U \gg \gamma_0$ (see text for details).
	    }
\end{figure}
Assuming the atom--atom interaction to be much stronger than the dipole--dipole coupling
(in essence, $U \gg \gamma_0$), 
we furthermore have (for each center-of-mass wavenumber $K$)
a bound state which we denote with the label $\nu=\mathrm{BS}$.
The relative wavefunction $\Psi^{(K, \mathrm{BS})}_{x \neq 0} = \alpha_K^{\abs{x}-1}$
is exponentially localized 
with respect to the relative coordinate
$x \equiv x_1-x_2$
($\alpha_K = -\ii \Gamma_1 \cos\left(K/2\right)/U$, $\abs{\alpha_K}<1$).
For $U \gg \gamma_0$,
this bound state is tightly confined,
\ie, $\Psi^{(K, \mathrm{BS})}_{x} = \delta_{\abs{x},1}$,
and describes a composite two-excitation object moving along the lattice.
Since only neighboring sites are occupied,
the minimal spatial separation
between two excitations is given by the lattice constant $a$.
The bound state's eigenenergy reads 
$E^{(2)}_{K, \mathrm{BS}} = 2 \omega_0 - \ii \Gamma_0 + \left( U^2 - \Gamma_1^2 \cos^2\left(K/2\right) \right)/U$
and contains an interaction-induced energy-shift. 
For $U \gg \gamma_0$, we have the expressions
$\mathrm{Re}(E^{(2)}_{K, \mathrm{BS}}) \simeq 2 \omega_0 + U$
and
$\mathrm{Im}(E^{(2)}_{K, \mathrm{BS}}) \simeq -\gamma_0$.
While two excitations approximately all decay at a rate of 
$\Gamma^{K \nu}_{\mathrm{tot}} \equiv -2 \mathrm{Im}(E^{(2)}_{K \nu}) \simeq 2\gamma_0$ 
for both $\nu=p$ and $\nu=\mathrm{BS}$, the bound states' energies 
are detached from the quasi-continuum 
of scattering states (see \figref{fig:disprelandsimp}).

Next, we formulate a Lindblad equation
$\partial_t \hat \varrho = \ii [ \hat \varrho, \hat{H}^\prime ] + \mathcal{L}(\hat \varrho)$
based on the single- and two-excitation eigenstates \cite{footnoteprocedure}.
Employing projectors $\hat{S}_{r;r \prime} \equiv \ketsmall{r} \brasmall{r^\prime}$,
the density matrix reads
$\hat \varrho = \sum_{r r^\prime} \varrho_{r; r^\prime} \hat{S}_{r; r^\prime}$
(where $r, r^\prime \in \{ 0, \{ k \}, \{ K \nu \} \}$).
For the coherent dynamics, we now utilize
$\hat{H}^\prime = \sum_{K \nu} \mathrm{Re}(E^{(2)}_{K \nu}) \hat{S}_{K \nu; K \nu} + \sum_{k} \mathrm{Re}(E^{(1)}_{k}) \hat{S}_{k; k}$.
The incoherent part
$\mathcal{L}(\hat \varrho) = \sum_s \left( R_s \hat \varrho R^{\dagger}_s - ( R^{\dagger}_s R_s \hat \varrho + \hat \varrho R^{\dagger}_s R_s )/2 \right)$
(with $s \in \{  \{K\nu;k\}, \{k;0\} \}$)
is constructed from the dissipators $R^{\dagger}_{K\nu;k} = \sqrt{\Gamma^{K \nu}_k} \hat{S}_{K \nu; k}$
and $R^{\dagger}_{k;0} = \sqrt{\Gamma^{k}_0} \hat{S}_{k;0}$.
Here, we introduced
$\Gamma^{K \nu}_k \equiv b^{(K \nu)}_k \cdot \Gamma^{K \nu}_{\mathrm{tot}} \simeq 2 \gamma_0 b^{(K \nu)}_k$,
where the branching ratio $b^{(K \nu)}_k$ accounts for the contribution
of the decay path $\ketsmall{K \nu} \rightarrow \ketsmall{k}$ to the overall decay of state $\ketsmall{K \nu}$.
The resulting equations of motion are given in the supplemental material \cite{supplement}.

In this paper, we focus on the emission properties of a bound state $\ketsmall{K, \mathrm{BS}}$
and, in a first step, study its spontaneous emission dynamics.
Assuming the system has been prepared in a pure eigenstate at time $t=0$,
the dynamics simplifies to 
\begin{eqnarray}
 \label{eq:solSE1}
 \varrho_{K, \mathrm{BS}; K, \mathrm{BS}} (t)
  &\simeq& \ee^{- 2 \gamma_0 t} \mathcomma \\
 \label{eq:solSE2}
 \varrho_{k; k} (t)
  &\simeq& 2 b^{(K, \mathrm{BS})}_k \left( \ee^{-\gamma_0 t}  -  \ee^{-2 \gamma_0 t}  \right)  
 \mathperiod
\end{eqnarray}
In order to relate these quantities to the scattered light in the far field, we 
write the Glauber decomposition of the electric field operator 
as \cite{glauber,mandelwolf}
$\hat{\vv{E}}^{(-)}(\vv{r},t) = \xi \vv{w}(\vv{r}) \sum_n  \sigma^{+}_{i}(t-t_n)$.
Here, $\xi = \omega_0^2 / (4 \pi \epsilon_0 c^2)$ ($\epsilon_0$ is the vacuum permittivity),
$\vv{w}(\vv{r}) = \left( \vv{d} - (\vv{d} \cdot \vv{r}) \vv{r} /r^2 \right) / r$ 
signifies the single-atom dipole field pattern ($\vv{d}$ is the dipole moment, 
$r \equiv \abssmall{\vv{r}}$),
and $t_n=\abssmall{\vv{r}-\vv{r}_n}/c \simeq r/c - \sin(\beta) n /c$ 
is the retarded time for a photon emitted
by atom~$n$.
The detector positioned at $\vv{r}$ is characterized by an observation angle 
$\beta$, where $\beta=0$ represents detection perpendicular to the atomic chain
(see also \figref{fig:sketchandredHS}).
Expanded in terms of eigenstates and exploiting
\cite{ficek} 
$\hat{S}_{k;0}(t-t_n) \simeq \exp(\ii \Delta^k_0 \sin(\beta) n / c) \hat{S}_{k;0}(t-r/c)$
as well as
$\hat{S}_{K \nu;k}(t-t_n) \simeq \exp(\ii \Delta^{K \nu}_k \sin(\beta) n / c) \hat{S}_{K \nu;k}(t-r/c)$,
the electric field operator takes the form
\begin{eqnarray}
 \label{eq:opefield}
 && \hat{\vv{E}}^{(-)}(\vv{r},t) =
 \xi \vv{w}(\vv{r}) \sqrt{M} \times 
 \\ \nonumber
 && ~~~ \sum_k 
 \Bigg{(} 
  \delta_{ k, \left[ \Delta^k_0 \sin(\beta)/c \right]_{\frac{2 \pi}{a}}} \hat{S}_{k;0}(t_{\mathrm{ret}})
  \\ \nonumber
  && ~~~~ ~+~ \sum_{K \nu} 
  \delta_{K-k , \left[ \Delta^{K \nu}_k \sin(\beta) / c \right]_{\frac{2 \pi}{a}}}
  \left( \bar{\eta}^{(K \nu)}_{\frac{K}{2}-k} \right)^* 
  \hat{S}_{K \nu;k}(t_{\mathrm{ret}})
 \Bigg{)}  
\mathcomma
\end{eqnarray}
where 
$\Delta^{K \nu}_k \equiv \mathrm{Re}(E^{(2)}_{K \nu}-E^{(1)}_k)$,
$\Delta^{k}_0 \equiv \mathrm{Re}(E^{(1)}_{k})$,
and $t_{\mathrm{ret}} \equiv t-r/c$.
In this expression, the Kronecker symbols are a consequence of the lattice sum over all 
atom positions
and determine the allowed emission angles ($\left[ l \right]_{m}$ signifies
$l$ modulo $m$).
To be precise, the decay of single-excitation states $\ketsmall{k} \rightarrow \ketsmall{0}$ 
requires 
$ka= 2\pi \left[ (a / \lambda_{\mathrm{at}}) \sin \beta \right]_{1}$
(using $\Delta^{k}_0 \simeq \omega_0$),
whereas for $\ketsmall{K, \mathrm{BS}} \rightarrow \ketsmall{k}$ we have
$(K-k)a= 2\pi \left[ (a / \lambda_{\mathrm{at}}) \sin \beta \right]_{1}$
(using $\Delta^{K, \mathrm{BS}}_k \simeq \omega_0 + U$ and assuming $\omega_0 \gg U$).
These expressions are reminiscent of Bragg's law and
the emission angles are determined by matching the wavenumbers
transferred to the free-space photon field.
For the remainder, $\bar{k}=\bar{k}(\vv{r})$ denotes the wavenumber that can be detected
at $\vv{r}$ (which allows us to set $k=\bar{k}$ in the first and $k=K-\bar{k}$ in the second sum
of \Eqref{eq:opefield}, respectively).
The quantity $\bar{\eta}^{(K \nu)}_{{K}/{2}-k}$ in \Eqref{eq:opefield} 
is of central importance in this paper and will be discussed later.

From \Eqref{eq:opefield}, we can now construct arbitrary far-field observables
such as the intensity that can be obtained for $\tau=0$ from
$\hat{G}^{(1)}(\vv{r},t,t+\tau) \equiv \hat{\vv{E}}^{(-)}(t) \hat{\vv{E}}^{(+)}(t+\tau)$
(where $\hat{\vv{E}}^{(+)} = ( \hat{\vv{E}}^{(-)} )^\dagger$).
In the context of spontaneous emission from a bound state (Eqs.~(\ref{eq:solSE1})--(\ref{eq:solSE2})), 
the expectation value 
$G^{(1)}(\vv{r},t) \equiv \expvalsmall{\hat{G}^{(1)}(\vv{r},t,t)} = \mathrm{tr} ( \hat{G}^{(1)}(\vv{r},t,t) \hat\varrho )$
reads
\begin{eqnarray}
\label{eq:G1BS}
\nonumber
\frac{   G^{(1)}(\vv{r},t)   }
      {   \xi^2 \abssmall{\vv{w(\vv{r})}}^2  M } 
     &=&  2 b^{(K, \mathrm{BS})}_{\bar{k}}
    \left( \ee^{-\gamma_0 t_{\mathrm{ret}}} - \ee^{-2 \gamma_0 t_{\mathrm{ret}}} \right) 
    \\ 
    && ~~~
    ~+~ 
    \abs{\bar{\eta}^{(K, \mathrm{BS})}_{\frac{K}{2}-\bar{k}}}^2
    \ee^{- 2 \gamma_0 t_{\mathrm{ret}}}
 \mathperiod
\end{eqnarray}

To finalize this result, we now turn to the discussion of the branching ratio 
$b^{(K,\mathrm{BS})}_k$ and the quantity $\abssmall{\bar{\eta}^{(K, \mathrm{BS})}_{K/2-\bar{k}}}^2$,
which are intimately linked to each other.
The contribution of the decay path $\ketsmall{K,\mathrm{BS}} \rightarrow \ketsmall{k}$ 
to the overall decay rate depends on the matrix element of the 
collective dipole moment operator \cite{haroche82}
$\hat{D}^- = (g^*_{\mu}/\sqrt{M}) \sum_{n=-N/2}^{N/2} \exp(-\ii \mu a n) \sigma^-_n $.
The wavenumber transferred to the photon field is $\mu=K-k$ and $g_{\mu}$ signifies
the atom--photon coupling (which can be considered being wavenumber-independent 
across the spectral window that is relevant here).
Explicit calculation of the matrix element gives
$\brasmall{k} D^- \ketsmall{K, \mathrm{BS}} = g^*_\mu \bar{\eta}^{(K, \mathrm{BS})}_{K/2-k}$,
where 
$\bar{\eta}^{(K, \mathrm{BS})}_{q} = (2/\sqrt{M}) \cos(qa)$.
As a result, we obtain the branching ratio via
$b^{(K,\mathrm{BS})}_k = \abssmall{   \brasmall{k} D^- \ketsmall{K, \mathrm{BS}}   }^2  /  \sum_{k^\prime} \abssmall{   \brasmall{k^\prime} D^- \ketsmall{K, \mathrm{BS}}   }^2=
(1/2) \abssmall{\bar{\eta}^{(K, \mathrm{BS})}_{K/2-k}}^2$.
However, 
the quantity $\abssmall{\bar{\eta}^{(K, \mathrm{BS})}_{K/2-k}}^2$ 
has yet another precise physical meaning.
Considering the Fourier transform of the two-body wavefunction, \ie, 
$(1/M) \sum_{n_1 n_2} \exp(-\ii k_1 a n_1) \exp(-\ii k_2 a n_2) \Phi^{(K \nu)}_{n_1 n_2} = (1/2) \bar{\eta}^{(K, \mathrm{BS})}_{(k_1-k_2)/2} \delta_{\left[K-k_1-k_2\right]_{2\pi/a},0}$,
shows that 
$\abssmall{\bar{\eta}^{(K, \mathrm{BS})}_{q}}^2$ 
is
the momentum distribution of 
the relative wavefunction.
For a bound state, this is a broad function in momentum space since the relative wavefunction
is tightly confined with respect to the relative coordinate.

\begin{figure}[t]
 \centering
 \includegraphics[width=0.48\textwidth]{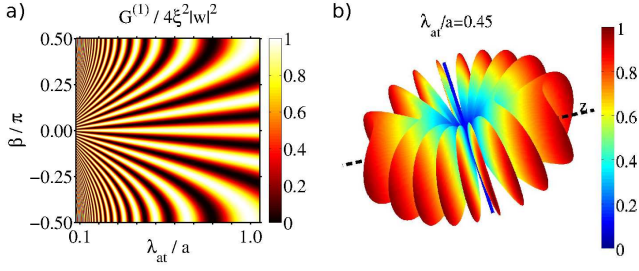}
  \caption{\label{fig:signature}
	    a)
	    Spontaneous emission pattern~$G^{(1)} / 4 \xi^2 \abssmall{\vv{w}}^2$
	    according to \Eqref{eq:G1BSfinal} 
	    (plotted for $t_{\mathrm{ret}}=0$ and $K=0$) 
	    as a function of the detection angle $\beta$
	    and the emission wavelength~$\lambda_{\mathrm{at}}$ 
	    (in units of~$a$).
	    As~$\lambda_{\mathrm{at}}/a$ increases, less Bragg orders become visible
	    but the width of the emission peaks increases.
	    b)~Angle dependence of the emitted radiation~$G^{(1)} r^2 / 4 \xi^2 \abssmall{\vv{d}}^2$.
	    The black dashed line indicates the direction of the atomic chain
	    and the blue line denotes the alignment of the dipole moments.
	    The pattern exhibits a toroidal-like structure 
	    (a remnant of the single-dipole pattern)
	    with lobes resulting from the properties of the bound state's 
	    momentum distribution.}
\end{figure}
Inserting 
these expressions 
into \Eqref{eq:G1BS}, we arrive at
\begin{equation}
\label{eq:G1BSfinal}
\frac{   G^{(1)}(\vv{r},t)   }
      {  4 \xi^2 \abssmall{\vv{w(\vv{r})}}^2 } 
     =   
    \cos^2\left( \frac{Ka}{2} - \frac{2 \pi a}{\lambda_{\mathrm{at}}} \left[ \sin \beta \right]_{\frac{\lambda_{\mathrm{at}}}{a}} \right)
    \ee^{- \gamma_0 t_{\mathrm{ret}}}
\mathcomma
\end{equation}
which reveals that the temporal decay of the intensity in the far field 
is mono-exponential (rate $\gamma_0$).
As this would also be observed for light emitted by a single-excitation state $\ketsmall{k}$
(where 
$G^{(1)}(\vv{r},t) / \abssmall{\vv{w(\vv{r})}}^2  \propto \exp\left( -\gamma_0 t_{\mathrm{ret}} \right) \delta_{k \bar{k}(\vv{r})}$),
the temporal decay as such cannot serve as an unambiguous signature for the
existence of a bound state.
However, the angle-dependent emission pattern as shown in \figref{fig:signature}
is a characteristic of the two-body bound state
whose peculiar features become most apparent when normalized
to the single-dipole pattern~$\abssmall{\vv{w(\vv{r})}}^2$.
Since the bound state sets a minimal spatial scale
(\ie, the minimal separation~$a$ between two excitations),
the momentum distribution covers a finite window in momentum space, 
which translates into 
emission peaks having a finite width (as a function of the emission angle).
This is in stark contrast to the sharp peaks that would be observed for 
a delocalized single-excitation state.

Admittedly, the preparation of a pure eigenstate $\ketsmall{K,\mathrm{BS}}$ may pose
severe challenges from a practical point of view.
We therefore now demostrate that the same characteristic far-field signature of the two-body 
bound state can also be obtained when the system is probed optically 
in a very simple way.
To this end, we envision a weak (\ie, strongly attenuated) 
incoherent driving field  
(\eg, pseudo-thermal light \cite{meschedebuch}) 
with a pump rate $\abssmall{\mathcal{P}}^2$ and
a spatial plane-wave pattern.
The projection of the external field's wavevector (magnitude $k_L$) on the atomic chain \anfz{imprints}
the wavenumber $k_P=\left[ k_L \sin\beta_{\mathrm{exc}} \right]_{2 \pi/a}$ (see \figref{fig:sketchandredHS})
and we assume $k_L = 2 \pi / \lambda_{\mathrm{at}}$.
A weak drive with $\Xi \equiv \abssmall{\mathcal{P}}^2 / \gamma_0 \ll 1$
allows us to work in the truncated Hilbert space of at most two excitations.
For an incoherent pump, the equations of motion reduce to a set of rate equations \cite{supplement},
yielding the steady-state occupation numbers (up to second order in $\Xi \ll 1$)
\begin{eqnarray}
 N_{K \nu} &\equiv& \varrho_{K \nu; K \nu} = 
 \frac{\Xi^2}{2} \abs{\bar{\eta}^{(2 k_P, \nu)}_{0}}^2 \delta_{K,2k_P}
 \mathcomma \\
 N_{k} &\equiv& \varrho_{k;k} = 
 \Xi \delta_{k k_P} + \Xi^2 \sum_{\nu} b^{(2 k_P, \nu)}_k \abs{\bar{\eta}^{(2 k_P, \nu)}_{0}}^2
 \mathperiod
\end{eqnarray}
The external pump only directly drives single-excitation states $\ketsmall{k=k_P}$ such that states
$\ketsmall{k \neq k_P}$ are populated via spontaneous emission of a two-excitation state 
$\ketsmall{2 k_P, \nu}$, which is of order $\Xi^2$.
Furthermore, only two-excitation states with $K=2k_P$ are excited (see \figref{fig:sketchandredHS}).

The emission spectrum emerging under these conditions is given by \cite{orszag}
${S}(\vv{r},\omega) =  2 \mathrm{Re}\left[ \int_{0}^{\infty} \mathrm{d} \tau \exp(\ii \omega \tau) \expvalsmall{ {G}^{(1)}(\vv{r},t,t+\tau) } \right]$,
which requires the calculation of the two-time correlation~${G}^{(1)}(\vv{r},t,t+\tau)$.
Employing the quantum regression theorem \cite{mandelwolf} leads to
\begin{eqnarray}
 \nonumber
 \frac{ \expvalsmall{ {G}^{(1)}(\vv{r},t,t+\tau) } }{\xi^2 \abs{\vv{w}(\vv{r})}^2 M} 
 &=& \ee^{-\ii \left( \omega_0 + U \right) \tau } \ee^{ - \frac{3}{2} \gamma_0 \tau } 
    \abs{ \bar{\eta}^{(2 k_P,\mathrm{BS})}_{k_P-\bar{k}} }^2
    N_{2 k_P, \mathrm{BS}}
 \\
 && ~~~~~ + \dots 
\mathperiod
\end{eqnarray}
Here, we have only specified the contribution from the scattered field 
that oscillates with the frequency $\omega_0 + U$
since we are now going to exploit the bound states' separation in energy from the 
band of scattering states (\cf \figref{fig:disprelandsimp}).
In other words, 
since $U \gg \gamma_0$, 
the corresponding emission spectrum 
at the frequency~$\omega_0+U$ has practically no overlap to
transitions around~$\omega_0$.
Specifically, measured at $\vv{r}$ (elevation $\beta$)
and, for convenience, normalized to the value recorded at a fixed direction $\vv{r}^\prime$
($\abssmall{\vv{r}^\prime} = \abssmall{\vv{r}}$, elevation $\beta_{\mathrm{exc}}$) in the $y$-$z$-plane,
we can write
\begin{eqnarray}
\label{eq:sigfinalgen}
 && 
 \frac{S\left( \vv{r} ,\omega = \omega_0 + U \right)}
      {S\left( \vv{r}^\prime ,\omega = \omega_0 + U \right)} 
 \cdot
 \frac{ \abssmall{\vv{d}}^2 / r^2 }{ \abssmall{\vv{w}(\vv{r})}^2 }
 \simeq
 \frac{\abs{ \bar{\eta}^{(2 k_P,\mathrm{BS})}_{k_P-\bar{k}} }^2}{\abs{ \bar{\eta}^{(2 k_P,\mathrm{BS})}_{0} }^2}
 \\ 
 \nonumber
 && ~~~~~~~~~~ = 
 \cos^2
 \left[ k_P a - \frac{2 \pi a}{\lambda_{\mathrm{at}}} \left[ \sin \beta \right]_{\frac{\lambda_{\mathrm{at}}}{a}}
 \right]
 \mathperiod
\end{eqnarray}
This is the same signature as obtained 
in the context of spontaneous emission from a pure eigenstate (see \Eqref{eq:G1BSfinal}),
even though here the external probing field is incoherent and weak.
Moreover, Eqs.~(\ref{eq:G1BSfinal}) and~(\ref{eq:sigfinalgen})
do not only represent an explicit far-field feature for the existence of a 
bound state on a lattice. 
These expressions can be utilized to extract the relative wavefunction's 
complete momentum distribution $\abssmall{\bar{\eta}^{(2k_P, \mathrm{BS})}_q}^2$.
This is achieved through tuning the argument $q=k_P(\beta_{\mathrm{exc}})-\bar{k}(\beta)$
across the first Brillouin zone by, for instance, varying the detection angle $\beta$ 
while keeping the excitation angle $\beta_{\mathrm{exc}}$ fixed.
The spectrum only needs to be recorded at a single frequency ($\omega = \omega_0 + U$)
and the presented scheme does not rely on single-atom addressability and/or manipulation 
techniques.

In conclusion,
we have analyzed the signatures that emerge from the excitation of a collective two-body
bound state on a lattice of atoms 
and found characteristic, angle-dependent far-field features in the scattered light.
For the future, 
we plan to utilize the generic theoretical approach presented in this work 
to also explore the radiative properties of scattering states 
and to analyze more sophisticated excitation and detection schemes.


\begin{thebibliography}{99}
\bibitem{footnoteprocedure}
For the construction of the Lindblad equation, 
we use the energies and total decay rates of 
the system's eigenstates as obtained from 
the non-Hermitian Hamiltonian~(\ref{eq:theH}) as an input,
and we work with the basis
corresponding to the non-dissipative Hamiltonian~$\hat{H}^\prime$
(realized by setting $\mathrm{Re}(\Gamma_0) = \mathrm{Re}(\Gamma_1) = 0$).
The branching ratio~$b^{(K,\mathrm{BS})}_k$ is calculated with respect to 
the non-dissipative basis.
\end{thebibliography}

\clearpage

{\bf Supplemental Material}
\\

The equations of motion resulting from a Lindblad equation 
as discussed in the main text are
\begin{eqnarray}
 \label{eq:eqomstart}
 \partial_t \varrho_{K \nu;K^\prime \nu^\prime} &=& 
 -  \frac{ \Gamma^{K \nu}_{\mathrm{tot}}  +  \Gamma^{K^\prime \nu^\prime}_{\mathrm{tot}} }{2} ~ \varrho_{K \nu; K^\prime \nu^\prime}
 + \mathcal{Q}_{K \nu; K^{\prime} \nu^{\prime}}
 \mathcomma \\
 \nonumber
 \partial_t \varrho_{k;k^\prime} &=& 
 - \frac{ \Gamma_{k} + \Gamma_{k^\prime} }{2} ~ \varrho_{k; k^\prime}
 \\ 
  && ~~~ + ~ \delta_{k, k^\prime} ~ \sum_{K \nu}
    \Gamma^{K \nu}_k \varrho_{K \nu; K \nu}
    + \mathcal{Q}_{k; k^{\prime}}
 \mathcomma \\
 \label{eq:eqomimed}
 \partial_t \varrho_{0;0} &=&
 \sum_k \Gamma_k \varrho_{k;k} 
 + \mathcal{Q}_{0; 0}
 \mathcomma \\
 \partial_t \varrho_{K \nu;k} &=&
 \left( -\ii \Delta^{K \nu}_k - \frac{\Gamma^{K \nu}_{\mathrm{tot}} + \Gamma_k}{2} \right) \varrho_{K\nu; k}
 \mathcomma \\
 \partial_t \varrho_{K \nu;0} &=&
 \left( -\ii \Delta^{K \nu}_0 - \frac{\Gamma^{K \nu}_{\mathrm{tot}}}{2} \right) \varrho_{K\nu; 0}
 \mathcomma \\
 \label{eq:eqomend}
 \partial_t \varrho_{k;0} &=&
 \left( -\ii \Delta^k_0 - \frac{\Gamma_k}{2} \right) \varrho_{k;0}
 \mathcomma
\end{eqnarray}
where 
$\Delta^{K \nu}_k \equiv \mathrm{Re}(E^{(2)}_{K \nu}-E^{(1)}_k)$,
$\Delta^{K \nu}_0 \equiv \mathrm{Re}(E^{(2)}_{K \nu})$,
and 
$\Delta^{k}_0 \equiv \mathrm{Re}(E^{(1)}_{k}) \simeq \omega_0$.
For a bound state ($\nu = \mathrm{BS}$) we have $\Delta^{K, \mathrm{BS}}_k \simeq \omega_0+U$
and $\Delta^{K, \mathrm{BS}}_0 \simeq 2\omega_0 + U$,
whereas for a scattering state ($\nu=p$) we get 
$\Delta^{Kp}_k \simeq \omega_0$ and $\Delta^{Kp}_0 \simeq 2\omega_0$.

In the context of spontaneous emission, 
the source terms $\mathcal{Q}_{r;r^{\prime}}$ are zero.  
In this case,
the solution to Eqs.~(\ref{eq:eqomstart})--(\ref{eq:eqomend}) 
reads 
\begin{eqnarray}
 \label{eq:sol1}
 \varrho_{K \nu; K^\prime \nu^\prime} (t)
  &\simeq& \ee^{- 2 \gamma_0 t} \varrho_{K \nu; K^\prime \nu^\prime}(0) \mathcomma \\
 \nonumber
 \label{eq:sol2}
 \varrho_{k; k^\prime} (t)
  &\simeq& \delta_{k k^\prime} \sum_{K \nu} 2 b^{(K \nu)}_k \left( \ee^{-\gamma_0 t}  -  \ee^{-2 \gamma_0 t}  \right) \varrho_{K \nu; K \nu}(0) \\
  && + ~ \ee^{- \gamma_0 t} \varrho_{k;k^\prime}(0) \mathcomma \\
  \label{eq:sol3}
 \varrho_{0; 0}(t)
  &=& 1 - \sum_{K \nu} \varrho_{K \nu; K \nu}(t) - \sum_k \varrho_{k; k}(t) \mathcomma \\
  \label{eq:sol4}
 \varrho_{K \nu; k}(t)
  &\simeq& \ee^{-\ii \Delta^{K \nu}_k t} \ee^{- 3 \gamma_0 t / 2 } \varrho_{K \nu; k}(0) \mathcomma \\
  \label{eq:sol5}
 \varrho_{K \nu; 0}(t) 
  &\simeq& \ee^{\ii \Delta^{K \nu}_0 t} \ee^{-\gamma_0 t} \varrho_{K \nu; 0}(0) \mathcomma \\
  \label{eq:sol6}
 \varrho_{k; 0}(t) 
  &\simeq& \ee^{-\ii \omega_0 t} \ee^{- \gamma_0 t / 2} \varrho_{k; 0}(0) 
 \mathperiod
\end{eqnarray}
Equations~(2) and~(3) in the paper result by choosing the initial condition of a pure bound state,
\ie, $\varrho_{K, \mathrm{BS}; K, \mathrm{BS}} (0) = 1$.

For an incoherent and weak driving field as described in the main text,
Eqs.~(\ref{eq:eqomstart})--(\ref{eq:eqomimed}) reduce to a set of rate equations by setting
\begin{eqnarray}
 Q_{K \nu; K \nu} &=& 
 \abssmall{\mathcal{P}}^2 \abs{\bar{\eta}^{(K \nu)}_{\frac{K}{2}-k_P}}^2 \varrho_{K-k_P;K-k_P}
 \mathcomma \\
 Q_{k; k} &=& 
 \delta_{k k_P} \abssmall{\mathcal{P}}^2 \varrho_{0;0}
 - \abssmall{\mathcal{P}}^2 \abs{\bar{\eta}^{(k+k_P, \nu)}_{\frac{1}{2} \left( k_P-k \right)}}^2 \varrho_{k;k}
 \mathcomma \\
 Q_{0; 0} &=& - \abssmall{\mathcal{P}}^2 \varrho_{0;0}
 \mathperiod
\end{eqnarray}
Equations~(7) and~(8) in the paper are the steady-state solution when
$\Xi=\abssmall{\mathcal{P}}^2/\gamma_0 \ll 1$.
Note that for $\Xi \ll \nolinebreak 1$, Eqs.~(\ref{eq:sol4})--(\ref{eq:sol6}) 
remain practically unaffected, which is exploited for the calculation
of the two-time field--field correlation function (Eq.~(9) in the paper).

\end{document}